\documentstyle [12pt]{article}

\def\H{{\cal H}}
\def\K{{\cal K}}

\def\n{{\cal N}}

\newcommand{\lbl}[1]{\label{eq: #1}}
\newcommand{\rf}[1]{\ref{eq: #1}}

\font\romdix=cmr10
\font\sldix=cmsl10
\font\bfdix=cmbx10
\newcommand{\bib}[3]{{\maj #1},\ {\romdix #2}.\ {\sldix #3}.}
\newcommand{\bibb}[4]{{\maj #1}\  {\romdix and}\ {\maj #2},\ {\romdix
#3}.\ {\sldix #4}.} 
\newcommand{\bibbb}[5]{{\maj #1},\, {\maj #2}\ {\romdix and}\
{\maj #3},\ {\romdix #4}.\ {\sldix #5}.}
\newcommand{\bibbbb}[6]{{\maj #1},\, {\maj #2},\, {\maj #3}\ {\romdix
and}\ {\maj #4},\ {\romdix #5}.\ {\sldix #6}.}

\font\maj=cmcsc10
\font\itdix=cmti10

\def\N{\mbox{I\hspace{-.15em}N}}

\def\Z{\mbox{Z\hspace{-.3em}Z}}

\def\R{{\rm I\hspace{-.15em}R}}

\def\C{\hspace{3pt}{\rm l\hspace{-.47em}C}}

\def\uk{\underline{\cal K}}
\def\uh{\underline{\cal H}}
\def\uv{\underline{U}}
\def\uu{\underline{U}}
\def\un{\underline{\cal N}}
\def\uun{\underline{\underline{\cal N}}}

\def\id{\mathop{\rm Id}\nolimits}

\newcommand\dpx[1]{\frac{\partial}{\partial#1}}
\newcommand\ddpx[1]{\frac{\partial^2}{\partial#1^2}}

\def\b{\begin{equation}} \def\e{\end{equation}}
\def\bd{\begin{displaystyle}} \def\ed{\end{displaystyle}} 
\def\ba{\begin{array}} \def\ea{\end{array}} 
\def\rw{\rightarrow}
\def\bee{\begin{enumerate}}
\def\eee{\end{enumerate}}
\newcommand\hs[1]{\hskip#1pt}
\def\bes{\begin{eqnarray*}}
\def\ees{\end{eqnarray*}}
\def\be{\begin{eqnarray}}
\def\ee{\end{eqnarray}}

\def\le{\langle}
\def\re{\rangle}
\def\dg{^{\dag}}

\begin{document}

\title{ Gupta-Bleuler quantization \\ for  minimally coupled
scalar fields \\ in de Sitter space}  
\author{J-P. Gazeau$^1$, J. Renaud$^1$\thanks{e-mail: renaud@ccr.jussieu.fr} , M.V.
Takook$^{1,2}$\thanks{e-mail: takook@ccr.jussieu.fr}}  \date{\today}

\maketitle
{\it
\centerline{$^1$ Laboratoire de Physique Th\'eorique de la Mati\`ere Condens\'ee}
\centerline{ Universit\'e Paris $7$ Denis-Diderot,75251 Paris Cedex 05, FRANCE}
\centerline{$^2$  Department of Physics, Razi University, Kermanshah, IRAN}
} 
\begin{abstract}
We present in this paper a fully covariant quantization of the
minimally-coupled massless field on de Sitter space, thanks to a new 
representation of the canonical commutation relations.
We thus obtain a formalism free of any infrared 
divergence.
Our method is based on a
rigorous group theoretical approach combined with a suitable adaptation
(Krein spaces) of the Wightman-G\"{a}rding axiomatic for massless fields
(Gupta-Bleuler scheme). We make explicit the correspondence between
unitary irreducible representations of the de Sitter group and the field
theory on de Sitter space-time.  The minimally-coupled massless field is 
associated with a representation which is the lowest term of the discrete 
series of unitary representations of the  de Sitter group. In spite of the
presence of negative norm modes in the theory, no negative energy can be
measured: expressions as 
$\le n_{k_1}n_{k_2}\ldots|T_{00}|n_{k_1}n_{k_2}\ldots\re$ are always positive.
\end{abstract} 

\section{Introduction}

We present in this paper a fully covariant construction of the minimally-coupled quantum field on
the de~Sitter space time. We specially emphasize the covariance aspect 
which should be understood in terms of the action of the de~Sitter group ${\rm SO}_0(1,4)$.
The starting point of the field construction is, within the context of 
canonical  quantization, the adoption of a new 
representation of the canonical commutation relations ({\bf ccr}). This 
field, of the Gupta-Bleuler type, is free of infra-red as well as 
ultra-violet divergences (our approach yields a covariant 
renormalization of the stress tensor).
The construction is of course {\em not} coordinate dependent. However, we shall make use of a specific
choice of global coordinates
through this paper, namely those ones which are called conformal, 
in order to make the construction explicit.

The last decades have seen considerable interest in the quantization of fields
in curved space-times of different types \cite{bd,ideux}. It is here not needed
to detail the various physical motivations among which inflationary scenarii and
quantum gravity take up a central position: see for instance a recent discussion by 
Lesgourgues, Polarski, and Starobinsky \cite{lps}. In this respect, the two de Sitter
spaces, namely de Sitter and anti-de Sitter, have been intensively studied,
owing to their constant curvature $H$ directly related to a non-zero
cosmological constant $\Lambda=3H^2$, and to the fact that both are of maximal
symmetry. Their respective isometry (or relativity) groups are indeed the
only two  deformations \cite{ln,bll}
of the Poincar\'e group. It was thus appealing to attempt a covariant
construction of quantum field theories on such space-times. We refer in
particular to a recent rigorous formulation of such a theory for the scalar
``massive" fields on de Sitter space  and its subsequent thermic
interpretation, (\cite{bgm} and references therein). Here we use quotes for the
term massive since, in both de Sitter relativities, the concept of mass does not
exist by itself as a conserved quantity. The mass is a  
galilean or minkowskian physical quantity. It can be measured through
well-established procedures by galilean or minkowskian observers. The expression
``massive" field in de Sitter space refers to an object which has a non ambiguous
massive limit as space-time becomes flat. 

Group representation theory and its Wigner interpretation in terms of
{\it elementary system} allow one to
control  this limiting process through contraction of group representation
\cite{mn,do}. 
As a matter of fact, ``massive" representations of the de~Sitter group, i.e.
representations which contract to massive representations of the Poincar\'e group,
are those of the
principal series of representations of the de Sitter  group SO$_{\rm
o}$(1,4), whereas massive representations of the  anti-de Sitter group are
those of the (holomorphic) discrete series of the
anti-de Sitter group SO$_{\rm o}$(2,3). 
We call massive fields the fields which transform under these
representations.  We here employ the standard
terminology of representation theory for semi-simple Lie group \cite{k}.

The situation is different for ``massless" field. A reasonable
physical requirement one has to impose on the concept of massless quantity is
light-cone propagation \cite{ffg}. It is thus natural to
call massless these representations of SO$_{\rm o}$(1,4) and SO$_{\rm o}$(2,3)
which naturally extend to the conformal group SO$_{\rm
o}$(2,4) \cite{affs}. Hence, associated de
Sitter fields really deserve the name of (conformally coupled) massless
fields.

Applying this criterium, one finds that the AdS scalar massless field
transforms under a representation which lies at the lower limit of the
(holomorphic) discrete series for the anti-de Sitter group SO$_{\rm o}$(2,3).
On the other hand, the de Sitter scalar massless field transforms under a
specific representation of the complementary series of SO$_{\rm o}$(1,4).  Once
massive and massless de~Sitter fields are well identified in terms of group
representation theory, corresponding covariant quantum field theories can be
envisaged along the lines proposed by Wightman and G\"{a}rding in their  seminal
paper \cite{wg}.

Within this context,  the specific case of massless spin-two
fields in de Sitter space, i.e.  de Sitterian gravitational fields in
their linear approximation, involves a
specific scalar field, called minimally coupled massless field. Although the
word massless is traditional in this case, this field is not really massless
in the above sense: the field equation is not conformally invariant and the
involved representation does not extend naturally to a representation of the
conformal group. Moreover this representation  has no Poincar\'e limit in the
sense we have proposed in the above. For various reasons, 
among which quantum gravity or more generally
quantum cosmology occupy a central position, a large amount of literature has been devoted to
the quantization problem for this field \cite{af,kg}. Here we wish to contribute to
this long quest toward a satisfactory mathematical setting. One of us  recently
gave \cite{dbr} a rigorous covariant treatment of the quantization problem for the
minimally-coupled massless field in $1+1$ dimensional de Sitter space-time.  It is
natural (and almost straightforward) to extend the method to the $3+1$ dimensional
case. 

Therefore we address in this paper the question of
constructing, for the minimally coupled massless equation in $3+1$ dimensional de
Sitter space, a quantum field theory which be covariant according to criteria
adapted from the original Wightman-G\"{a}rding paper.
In this sense, a quantum field is, roughly speaking,  a distribution  $\varphi$
on space-time, solution of the field equation, with values in a set of
symmetric operators in some inner product  space and verifying some physically
reasonable properties. (In the following, we
underline the second-quantized spaces and the corresponding group action, in
order to make a clear distinction between the states and the elements of the
one-particle sector.)
   \begin{itemize}  \item{\bf[Covariance]} There exists a
unitary  representation $\uu$ of de~Sitter group on the space of states, and the
field is covariant 
  $$  \uu_g \varphi(x)\uu^{-1}_g=\varphi(g\cdot x)$$
for any $g$ in de~Sitter group and $x$ in space-time.
 \item{\bf[Existence of the vacuum]}
  There exists   a normalized state $|0\re$ called the
vacuum, which is invariant under the representation $\uu$
     $$\uu_g|0\re=|0\re\hs{10} \mbox{  for all  }  g\in G, $$
and which is {\it unique, up to physical equivalence}.
\item{\bf[Causality]} A local commutativity property holds:
   $$[\varphi(x_1),\varphi(x_2)]=0$$
as far as the points $x_1$ and $x_2$  are not
causally connected.\end{itemize}

This is usually achieved through a  Fock construction based upon  a suitable
one-particle sector Hilbert space. However Allen \cite{a} has proved that such a field
does not exist on de~Sitter space-time. The crucial point in Allen's proof
is the following. Any set of normalisable modes, complete in the sense that they, together with the
conjugate modes, span the constant functions, cannot be de~Sitter invariant. The
existence of a constant solution is at the origin of the problem. Allen has
noted that this constant function can be considered as the real part of a so-called zero
mode. This zero mode has positive norm, but it cannot be 
part of the Hilbertian structure of the one particle sector. Indeed, the action of the de~Sitter group
on this mode generates all the negative frequency solutions (with 
respect to the conformal time) of the field 
equation. Other approaches based on two-point functions (see \cite{l} 
for instance) also failed in 
this case. Indeed, for the same reason, there  is no 
covariant two-point function $G_{1}$´ of the positive type, i.e. 
obeying 
$$\int G_{1}(x,x')u^{*}(x)u(x')dx\,dx'\geq 0$$
 for any solution $u$.´´
The
usual procedure for overcoming this difficulty is to adopt a restrictive version of covariance by considering fields
which are covariant with respect to a subgroup of the de~Sitter group: this is the
so-called ``symmetry breaking''.

Our approach is different. We instead require full covariance as well as
causality, but the Allen's statement shows that we have to give up something. Before
discussing this point, let us remark that there is a deep analogy between
the zero mode problem and the quantization of the electromagnetic field.
The Lagrangian of the minimally coupled field 
$${\cal L}=\sqrt{|g|}\partial_\mu\phi\partial^\mu\phi$$
is invariant under the global transformation $\phi\mapsto\phi+\lambda$ which is
like a gauge transformation. As it is well known, the correct procedure for quantizing
electromagnetism does not consist in weakening  the covariance. One has to adopt the Gupta-Bleuler
quantization, and this is precisely what we  do here for the minimally coupled field. 

At this point, let us precise what we mean by Gupta-Bleuler formalism. 
In  electrodynamics the Gupta-Bleuler triplet
$V_g\subset V\subset V'$ is defined as follows \cite{bfh}\cite{jpg}.
The space $V_g$ is the space of longitudinal photon states or ``gauge
states'', the space $V$ is the space of positive frequency solutions
of the field equation verifying the Lorentz condition, and $V'$ is 
the space of all positive frequency  solutions of the field equation,
containing non-physical states. The Klein-Gordon inner product defines
an indefinite inner product on $V'$ which is Poincar\'e invariant. All three 
spaces carry representations of the Poincar\'e group but $V_g$ and $V$ are not
invariantly complemented. The quotient space $V/V_g$ of states up to a gauge 
transformation is the  space of physical one-photon states. The
quantized field acts on the Fock space built on $V'$, which is not a
Hilbert space, but is instead an indefinite inner product space. 

We proceed in a similar manner for the minimally coupled field. The set $\n$ of
constant functions will play the role of $V_g$. We also obtain a physical space
$\K$ carrying a unitary representation of the de~Sitter group. However, this space is
{\it not} a Hilbert space: the Klein-Gordon inner product is degenerate (although
positive), and there is of course no contradiction with the Allen's result.
Moreover the representation of the de~Sitter group is not irreducible (although
indecomposable). As
discussed at length in \cite{dbr}, the field must be written on  a nondegenerate inner product
space. As a consequence we must introduce as a total space $\H$ a
much larger space. The latter contains 
auxiliary states which can be of negative norm for the usual Klein-Gordon inner
product. Nevertheless this does not mean that negative energies could be
attainable in terms of observable measurements. Indeed, expressions 
like  $\le
n_{k_1}n_{k_2}\ldots|T_{00}|n_{k_1}n_{k_2}\ldots\re$ are  positive 
for any physical state $|n_{k_1}n_{k_2}\ldots\re$. Moreover
this construction yields an automatic and covariant renormalization of the stress tensor: the
above expression is free of any infinite term. This clearly indicates the crucial role 
played by the negative modes: they allow one to overcome in a totally covariant way
the zero mode problem. 

Again, we emphasize the fact that our minimally coupled field is defined on a space which
 is {\em not} a Hilbertian Fock space, and there is no contradiction with the
result of Allen. This is due to the fact that the one-particle sector itself is not
a Hilbert space (the inner product is not positive). The physical space {\it stricto sensu} is the
quotient space $\K/\n$. This is a Hilbert space carrying a unitary irreducible representation of
the de~Sitter group. Nevertheless, such a quotient space is an abstract space and any attempt to
realize it as a space of solutions of the field equation requires to invert the above quotient map.
There are many ways to do this. None is natural ({\it i.e.} 
covariant). 
Any naive approach relies on such a construction (explicitly or not), and the
consequence is a symmetry breaking in the theory. 
A frequently adopted manner to achieve this unnatural implementation of
the Hilbert space structure in the theory is to write down the
massive theory and then to put $(m_H^2+\xi R)\rw0$ in (\rf{intact}). Infinite
divergences appear in this computation, from which it is often claimed that  the
vacuum state is not normalisable. On the contrary, our approach is to start from the
minimally coupled framework (equation and its set of solutions), and no divergence exists. Indeed, all the states
are of finite norm with respect to the natural inner product (see (\rf{kgie})
below): in particular the norm of a global gauge state vanishes and 
no infra-red divergence appears.

 We shall first present, in the next section, the de Sitter machinery. By this
we mean a set of  definitions and notations concerning geometry and wave
equations on one hand, and the relevant group-theoretical material on the other
hand. We shall especially insist on the terminology in use in representation
theory in order to make the reader more familiar with a complete
classification of unitary irreducible representations of SO$_{\rm o}$(1,4) and
the respective physical meaning of the latter. Section 3 is devoted to the
description of the space of solutions of involved scalar wave equations. In
Section 4, we make explicit  the
Gupta-Bleuler structure lying behind the minimally-coupled massless field.
Indeed, the interesting one-particle sector in the space of solutions of
$\Box\phi=0$ in de Sitter can be structured into a so-called Krein space =
Hilbert $\oplus$ anti-Hilbert, of which an invariant subspace is made of 
 constant functions. 
 In section 5 we present the new representation of the {\bf ccr} from
 which the quantum field is obtained. The Fock space carrying this representation is 
 based on the  Krein space. In order to control to what extent
our quantization scheme is physically well-founded, we compute in 
section 6 the mean
values of the stress tensor in our vacuum (we find zero!), and in excited
states (we find positive values, as it should be reasonably expected even though
the representation of the canonical commutation relations ({\bf ccr}) involves 
negative norm solutions in order to
preserve de Sitter covariance). After a brief comment on the 
extension of our method to massive fields (section 7), we finally conclude 
in section 8.

\section{Presentation of the de Sitter machinery}

The de Sitter space is conveniently seen as a hyperboloid embedded in a
five-dimensional Minkowski space
         $$ M_H=\{ X\in \R^5|\;\;X^2=\eta_{\alpha\beta}X^\alpha
X^\beta=X_\alpha
       X^\alpha =-H^{-2}\},  $$
where $ \eta^{\alpha\beta} =\mbox{diag}(1,-1,-1,-1,-1)$.  The
(pseudo-)sphere $M_H$ is obviously invariant under five-dimensional
Lorentz transformation. Therefore de Sitter space has a ten-parameter
group of isometries, the de Sitter group O$(1,4)$. We only consider the
 connected component of the identity SO$_{\rm o}(1,4)$. We
are in particular interested by the Poincar\'e limit of the latter through
the group contraction $H\rw0$, i.e. when the curvature tends toward 0. The
ten infinitesimal generators $M_{\alpha\beta}$ in some unitary
representation of the de Sitter group obey the following well-known
commutation rules (with $\hbar=1)$ 
     \b\lbl{rc}[M_{\alpha\beta},M_{\gamma\delta}]=-i(\eta_{\alpha\gamma}
     M_{\beta\delta}-\eta_{\alpha\delta}M_{\beta\gamma}+\eta_{\beta\delta}
        M_{\alpha\gamma}-\eta_{\beta\gamma}M_{\alpha\delta}) .\e 

In this work, we shall make use of a system of bounded global coordinates
$(x^\mu,\;\mu=0,1,2,3)$ well-suited to describe a compactified version of dS,
namely S$^3 \times{\rm S}^1$ (Lie sphere). This system is given by
     $$\left\{
\ba{rcl}
X^0&=&H^{-1}\tan \rho\\
X^1&=&(H\cos\rho)^{-1}\,(\sin\alpha\,\sin\theta\,\cos\phi),\\
X^2&=&(H\cos\rho)^{-1}\,(\sin\alpha\,\sin\theta\,\sin\phi),\\
X^3&=&(H\cos\rho)^{-1}\,(\sin\alpha\,\cos\theta),\\
X^4&=&(H\cos\rho)^{-1}\,(\cos\alpha),\ea\right.$$
where $-\pi/2<\rho<\pi/2$, $0\leq\alpha\leq\pi$, $0\leq\theta\leq\pi$ and
$0\leq\phi<2\pi$.
 The coordinate $\rho$ is timelike and plays the role of a conformal 
 time. The closure of the $\rho$-interval is actually
involved when dealing with conformal action on compactified space-time.
The  de~Sitter metrics now reads
   \b ds^2=g_{\mu\nu}dx^\mu dx^\nu=\frac{1}{H^2 \cos^2\rho}
(d\rho^2-d\alpha^2-\sin^2\alpha\, d\theta^2-\sin^2\alpha\sin^2\theta
\,d\phi^2).\e

In the scalar representation carried by functions on $M_H$, the infinitesimal
generators (\rf{rc})  are given by \cite{ct}:
$$M_{ab}=
-i\left(X_b\frac{\partial}{\partial X_a}-X_a\frac{\partial}{\partial X_b}\right)\ 
a,b=0,1,2,3,4.$$
With our choice of coordinates the six generators of the compact $SO(4)$
subgroup, contracting to the Lorentz subalgebra when $H\rw0$,   read as follows. 
\bes
M_{12}&=&-i\dpx{\phi},\\
M_{32}&=&-i(\sin\phi\dpx{\theta}+\cot\theta\cos\phi\dpx{\phi}),\\
M_{31}&=&-i(\cos\phi\dpx{\theta}+\cot\theta\sin\phi\dpx{\phi}),\\
M_{41}&=&-i(\sin\theta\cos\phi\dpx{\alpha}
+\cot\alpha\cos\theta\cos\phi\dpx{\theta}
-\cot\alpha\frac{\sin\phi}{\sin\theta}\dpx{\phi}),\\
M_{42}&=&-i(\sin\theta\sin\phi\dpx{\alpha}
+\cot\alpha\cos\theta\sin\phi\dpx{\theta}
+\cot\alpha\frac{\cos\phi}{\sin\theta}\dpx{\phi}),\\
M_{43}&=&-i(\cos\theta\dpx{\alpha}
-\cot\alpha\sin\theta\dpx{\theta}).
\ees
The four generators contracting  to the space-time translations
 when $H\rw0$ read as follows.
\bes
M_{01}&=&-i(\cos\rho\sin\alpha\sin\theta\cos\phi\dpx{\rho}
+\sin\rho\cos\alpha\sin\theta\cos\phi\dpx{\alpha}
+\frac{\sin\rho\cos\theta\cos\phi}{\sin\alpha}\dpx{\theta}\\
&&\hskip330pt-\frac{\sin\rho\sin\phi}{\sin\alpha\sin\theta}\dpx{\phi}),\\
M_{02}&=&-i(\cos\rho\sin\alpha\sin\theta\sin\phi\dpx{\rho}
+\sin\rho\cos\alpha\sin\theta\sin\phi\dpx{\alpha}
+\frac{\sin\rho\cos\theta\sin\phi}{\sin\alpha}\dpx{\theta}\\
&&\hskip330pt+\frac{\sin\rho\cos\phi}{\sin\alpha\sin\theta}\dpx{\phi}),\\
M_{03}&=&-i(\cos\rho\sin\alpha\cos\theta\dpx{\rho}
+\sin\rho\cos\alpha\cos\theta\dpx{\alpha}
-\frac{\sin\rho\sin\theta}{\sin\alpha}\dpx{\theta}),\\
M_{04}&=&-i(\cos\rho\cos\alpha\dpx{\rho}
-\sin\rho\sin\alpha\dpx{\alpha}).
\ees

The O$(1,4)$-invariant measure on $M_H$ is 
   \b  d\mu= \sqrt{-g}\,dx^0dx^1 dx^2 dx^3=(\cos\rho)^{-4}\, d\rho\, d\Omega,\e
where $d\Omega=\sin^2\alpha \sin\theta\,d\alpha\,d\theta\,d\phi$ is the
O(4)-invariant measure on S$^3$. The Laplace-Beltrami operator on de~Sitter space
is given by 
$$ææ\Box=\frac{1}{\sqrt g}\partial _\nu\sqrt gg^{\nu\mu}\partial _\mu=
   H^2  \cos^4 \rho \frac{\partial }{\partial \rho}(\cos^{-2} \rho
    \frac{\partial }{\partial \rho})-H^2\cos^2 \rho\, \Delta_{3},$$ 
where 
$$ \Delta_{3}=\ddpx{\alpha}+2\cot\alpha\dpx{\alpha}+\frac{1}{\sin^2\alpha}
\ddpx{\theta}+\cot\theta\frac{1}{\sin^2\alpha}\dpx{\theta}
+\frac{1}{\sin^2\alpha\sin^2\theta}\ddpx{\phi}$$
 is the Laplace operator on the
hypersphere ${\rm S}^3$.

 The wave equation for scalar fields $\phi$ propagating on de Sitter
space can be derived from variational principle on the action integral
$(\hbar=1)$
     \b\lbl{intact} S(\phi)=\frac{1}{2}\int_{M_H}[g^{\mu\nu}\partial _{\mu}\phi
    \partial _{\nu}\phi-(m_H^2+\xi R)\phi^2]d\mu,\e
where $m_H$ is a ``mass", $R=12H^2$ is the Ricci
(or curvature) scalar, and $\xi$ is a positive gravitational
coupling with the de Sitter background. The variational principle applied to
(\rf{intact})  leads to the field equation 
    \b\lbl{scafiel} [\Box_H +(m_H^2 + \xi R)]\phi(x)=0.\e

The Klein-Gordon inner-product is defined for any $\phi,\psi$ solutions of
(\rf{scafiel}) by
 \b\lbl{kgie} \le\phi,\psi\re=i\int_\Sigma \phi^{*}´(\overrightarrow{
   \partial}_{\mu}-\overleftarrow{\partial}_{\mu})\psi d\sigma^\mu\equiv
i\int_\Sigma \phi^*\stackrel{\leftrightarrow}{
   \partial}_{\mu}\psi d\sigma^\mu,\e
where $\Sigma$ is a Cauchy surface, {\it i.e.} a space-like surface such that the Cauchy
data on $\Sigma$ define uniquely a solution of (\rf{scafiel}), and $d \sigma^\mu$
is the area element vector       on $\Sigma$. This product is de~Sitter invariant and
independent of the choice of $\Sigma$.
In accordance with our choice of global coordinate
system, the Klein-Gordon inner product (\rf{kgie}) reads
\renewcommand{\arraystretch}{0.6}
 \b \le\phi,\psi\re=\frac{i}{H^2}\int_{\rho=0}
 \phi^{*}´(\rho,\Omega)\stackrel{\leftrightarrow}{
\partial}_{\rho}\psi(\rho,\Omega)
d\Omega,\e  \renewcommand{\arraystretch}{1}
where $d\Omega=\sin^2\alpha\sin\theta\,d\alpha\,d\theta\,d\phi$ is
the invariant measure on S$^3$.

As we now explain, the equation (\rf{scafiel}) has a clear
group-theoretical content.
 Let us recall that the Casimir
operator $Q_0$ is defined by
\b    Q_0=-\frac{1}{2} M^{\alpha\beta}M_{\alpha\beta}.\e
This operator commutes with the action of the group
generators and, as a consequence, it is constant on each unitary
irreducible representation (UIR). As a matter of fact, the scalar UIR's can be
classified using the eigenvalues of $Q_0$. This allows to identify the
scalar UIR associated to each scalar field on de~Sitter space time because
  the Laplace-Beltrami operator  and the  Casimir operator are
proportional:
   $$ \Box_H=-H^2Q_0.$$
Rewritten in these terms, (\rf{scafiel}) reads
  $$ Q_0\phi=\kappa\phi,$$
with
  \b\lbl{kappa} \kappa= <Q_0>=(\frac{m_H}{H})^2 +12 \xi.\e

We consider only the positive values of $\kappa$ and we denote by $U^{(\kappa)}$
the scalar UIR corresponding to  the value $\kappa$ of the Casimir
operator. The classification of these  scalar representations is the
following~\cite{d}.  \bee \item For 
$$\kappa=\le Q_0\re\in[\frac{9}{4},+\infty[,$$
the corresponding UIR's $U^{(\kappa)}$ are known as elements of the
principal series of representations. They are written $\nu_{0,\kappa-2}$ in
\cite{d}.
 \item For
$$\kappa=\le Q_0\re\in]0,\frac{9}{4}[,$$
the corresponding UIR's $U^{(\kappa)}$ are known as elements of the complementary
series of representations. They are written $\nu_{0,\kappa-2}$ in
\cite{d}.
\item For 
$$\kappa=\le Q_0\re=0,$$
the corresponding  UIR $U^{(0)}$ is known as the first term of the scalar
discrete series of representations. It is written $\pi_{1,0}$ in \cite{d}.
\eee

The physical content of each one from the point of view of a minkowskian
observer (limit $H=0$) is the following.
\begin{itemize}
\item First, we consider the ``massive'' case, {\it i.e.} the values of
$\le Q_0\re$ corresponding to the principal series of
representation. In order to obtain the contraction of group
representations, the number $\kappa$, by which representations are
labelled, goes to infinity in such a way that $H^2\kappa\rw m^{2}´$. It has been
proven that the principal series UIR's
$U^{(\kappa)},\;\kappa\geq0,$ contracts toward the direct sum 
of two massive Poincar\'e UIR's ${\cal P}(\pm m)$ with negative and
positive energies respectively \cite{mn}:
\renewcommand{\arraystretch}{0.6}
 \b\lbl{contr} U^{(\kappa)} \stackrel{\ba{c} \kappa \rightarrow\infty\\
 H \rightarrow
0\ea}{\longrightarrow} {\cal P}(-m)\oplus
              {\cal P}(m). \e
\renewcommand{\arraystretch}{1} Note that the constraint $H^2\kappa\rw 
m^{2}´$
and the equation~(\rf{kappa}) imply that the quantity $m_H$, supposed to
depend on $H$, tends to the classical mass $m$ when the curvature goes to
zero.

\item Second we consider the massless case. As explained in the
introduction, we select the representation having a natural extension
to the conformal group. This representation is  $U^{(2)}$, an element of 
 the complementary series  (this corresponds to
$m_H=0$ and $\xi=1/6$). The representation $U^{(2)}$  extends to an   UIR  ${\cal
C}^+_0$ of the conformal group SO$_0(2,4)$ \cite{bb}. In contrast to the massive
case, the contraction process involves only one representation. The representation
involved for each value of $H$, including $H=0$ is equivalent to ${\cal
C}^+_0\oplus{\cal C}^-_0$ \cite{affs}.
 The following diagram
illustrates these connections:
   \b   \begin{array}{ccccccc}
  {\cal C}^-_0   &   \oplus        &{\cal C}^+_0  &
\stackrel{H\rw0}{\longrightarrow}
  &{\cal C}^-_0  
&\oplus &{\cal C}^+_{0}´\\ 
 &  &\bigsqcup 
&     &  \bigsqcup        & &\bigsqcup     
\\
     &             & U^{(2)}   & \stackrel{H\rw0}{\longrightarrow}&{\cal 
     P}^-(0)
&\oplus &{\cal P}^+(0)\\
    \end{array} \e
where the symbol $\bigsqcup$ means that the upper representation is an
extension of the lower one and ${\cal P}^\pm(0)$ are the
massless Poincar\'e UIR's with positive and negative energies
respectively. 

\item The minimally coupled field ($\le Q_0\re=0$) has no
minkowskian counterpart but it is interesting partly because this
field appears when treating the spin-two field. The involved UIR is
the first term $U^{(0)}$ of the discrete series of representations .
\end{itemize}

\section{Space of solutions }\label{solutions} 

Equation (\rf{scafiel}) can be solved by separation of variable
\cite{ct,kg}. We put
     $$ \phi(x)=\chi(\rho)D(\Omega), $$
where $\Omega\in{\rm S}^3$,
and obtain 
      \begin{eqnarray} [\Delta_{3}+C]D(\Omega)&=&0,\lbl{eqang}  \\
 (\cos^4 \rho \frac{d}{d\rho}\cos^{-2} \rho \frac{d}{d\rho}
          +C\cos^2 \rho
+(\frac{m_H}{H})^2+12\xi)\chi(\rho)&=&0.\lbl{eqrad}\end{eqnarray}

We begin with the angular part problem (\rf{eqang}). For
$C=L(L+2)$, $L\in\N$ we find the hyperspherical harmonics 
$D={\cal Y}_{Llm}$  which are defined by
$${\cal Y}_{Llm}(\Omega)
=\left(\frac{(L+1)(2l+1)(L-l)!}{2\pi^2(L+l+1)!}\right)^{\frac{1}{2}}
2^ll!\left(\sin\alpha\right)^lC_{L-l}^{l+1}\left(\cos\alpha\right)
Y_{lm}(\theta,\phi),$$
for $(L,l,m)\in\N\times\N\times\Z$ with $0\leq l\leq L$ and $-l\leq
m\leq l$. In this equation the $C_n^\lambda$ are Gegenbauer
polynomials \cite{t} and $Y_{lm}$ are ordinary spherical harmonics:
$$Y_{lm}(\theta,\phi)=
(-1)^m\left(\frac{(l-m)!}{(l+m)!}\right)^{\frac{1}{2}}
P_l^m(\cos\theta)e^{im\phi},$$
where $P_l^m$ are the associated Legendre functions.
With this choice of constant factors, the ${\cal Y}_{Llm}$'s obey the
orthogonality (and normalization) conditions:
$$\int_{\rm
S^3}{\cal Y}_{Llm}^{*}´(\Omega){\cal Y}_{L'l'm'}(\Omega)\,d\Omega
=\delta_{LL'}\delta_{ll'}\delta_{mm'}\ .$$

We now come to the radial-part problem (\rf{eqrad}). Let $\lambda$ be
defined by
\begin{eqnarray}
\lambda&=&\sqrt{\frac{9}{4}-\kappa}\mbox{ when 
}\frac{9}{4}\geq\kappa\geq0,\nonumber\\
\lbl{lambda}\lambda&=&i\sqrt{\kappa-\frac{9}{4}}\mbox{ when
}\frac{9}{4}\leq\kappa.\end{eqnarray}
 Following \cite{kg}, we obtain the solutions 
  \b\lbl{rad} \chi_{\lambda L}(\rho)=A_L(\cos
      \rho)^{\frac{3}{2}}[P^{\lambda}_{L+\frac{1}{2}}(\sin\rho)-
\frac{2i}{\pi}
    Q^{\lambda}_{L+\frac{1}{2}}(\sin\rho)].\e
Here $P^\lambda_n$ and $Q^\lambda_n$ are the Legendre functions on the
cut, and $A_L$ is given by
    $$ A_L=H\frac{\sqrt\pi}{2}\left(\frac{\Gamma(L-\lambda+\frac{3}{2})}
    {\Gamma(L+\lambda+\frac{3}{2})}\right)^{\frac{1}{2}}.$$
We then obtain the complete set of modes
\b\lbl{mode}\phi_{Llm}^\lambda(x)=\chi_{\lambda L}(\rho){\cal
Y}_{Llm}(\Omega),\  x=(\rho,\Omega)\in M_H,\e 
for the field equation $(\Box+\kappa)\phi=0$, where $\lambda$ is defined through
(\rf{kappa}) and (\rf{lambda}), except for the minimally-coupled field $\kappa=0$,
$\lambda=3/2$, for which the formulas break down.
Note that this family of modes verify the orthogonality prescription:
$$\le\phi_{L'l'm'}^\lambda,\phi_{Llm}^\lambda\re
=\delta_{LL'}\delta_{ll'}\delta_{mm'}\mbox{ and }
\le\phi_{L'l'm'}^\lambda,\left(\phi_{Llm}^\lambda\right)^{*}´\re=0.$$
This family can be used  to define the euclidean vacuum in the standard terminology.

We now turn our attention to the singular case $\lambda=\frac{3}{2}$
corresponding to the minimally coupled field. For $L\neq0$, we obtain the modes
$\phi^{\frac{3}{2}}_{Llm}$ that we write  $\phi_{Llm}$ for simplicity:
\b\lbl{modemn}\phi_{Llm}(x)=\chi_{L}(\rho)Y_{Llm}(\Omega),\e with
\b \chi_{L}(\rho)=\frac{H}{2}[2(L+2)(L+1)L]^{-\frac{1}{2}}
  \left(L e^{-i(L+2)\rho}+(L+2)e^{-iL\rho}\right).\e
 The normalization
constant $A_L$ breaks down at $L=0$. This is the famous ``zero-mode" problem.
The space generated by the $\phi_{Llm}$ for $L\neq0$ is not a complete set of
modes. 
Moreover this set is not invariant under the action of the de~Sitter group.
Actually, an explicit computation gives 
\b\lbl{ncov}
(M_{03}+iM_{04})\phi_{1,0,0}=-i\frac{4}{\sqrt{6}}\phi_{2,1,0}+\phi_{2,0,0}+
\frac{3H}{4\pi\sqrt{6}},\e
and the invariance is broken owing to the last term. As a consequence, canonical
quantization applied to this set of modes yields a non covariant field,
and this  is due to  the apparition  of the  last term in (\rf{ncov}).
Constant functions are of course solutions to the field equation.
So one is led to deal with the space generated by the $\phi_{Llm}$'s 
and by a constant function denoted here by $\psi_g$, this is 
interpreted as a 
gauge state as announced in the introduction.
This space, which {\it is invariant under the de~Sitter group}, is the space of
physical states as explained below. However, as an inner-product space
equipped with the Klein-Gordon inner product, it is a degenerate space
because the state $\psi_g$ is orthogonal to the whole space including
itself. Due to this degeneracy, canonical quantization applied to this set
of modes yields once again a non covariant field (see \cite{dbr} for a
detailed discussion of this fact).

Actually, for $L=C=\kappa=0$, the equation (\rf{eqrad}) is easily solved. 
We obtain two independent solutions of the field equation, including the constant
function discussed above:
 $$\psi_g=\frac{H}{2\pi}\mbox{ and }
\psi_s=-i\frac{H}{2\pi}[\rho+\frac{1}{2}\sin 2\rho ].$$
Note that the constants of normalization are chosen in order to have
$\le\psi_g,\psi_s\re=1$. 
One can now defines $\phi_{000}=\psi_g+\psi_s/2$. This is the ``true zero mode''
of Allen. We write  $\phi_{000}=\phi_{0}$ in the following´´. 
With this mode, one obtains a complete set of strictly positive norm
modes $\phi_{Lml}$ for $L\geq0$, but the space generated by these modes {\it is not
de~Sitter invariant}. For instance, we have
\b\lbl{ncovd}(M_{03}+iM_{04})\phi_{0}=(M_{03}+iM_{04})\psi_s
=-i\frac{\sqrt{6}}{4}\phi_{1,0,0}+-i\frac{\sqrt{6}}{4}\phi_{1,0,0}^{*}´
-\frac{\sqrt{6}}{4}\phi_{1,1,0}-\frac{\sqrt{6}}{4}\phi_{1,0,0}^{*}.\e As a
consequence the field obtained through canonical quantization and the usual 
representation of the {\bf ccr} from the set of
modes $\phi_{Lml}$ for $L\geq0$  is {\it not covariant}. Nevertheless, 
the above space is O(4) invariant, and with this set of modes one 
obtains by the usual construction a O(4)-covariant quantum field 
\cite{a}. 
On the other hand, although our covariant and causal field is also obtained by canonical quantization:
$$\varphi(x)=\sum_{k}\phi_k(x)A_k+\phi_k^*(x)A\dg_k,$$
whith $A_k$ and $A_k^{\dag}$  satisfying the {\bf ccr}:
$$[A_k,A\dg_{k'}]=2\delta_{kk'},\hs{10}
[A_k,A_{k'}]=0,\hs{10}[A\dg_k,A\dg_{k'}]=0,$$
we use a different  representation of the {\bf ccr}  in 
order to obtain the field as an operator valued distribution.

Note the
appearance of negative norm modes in (\rf{ncovd}) which is the price 
to pay in order to obtain
a fully covariant theory. The existence of these non physical states naturally 
leads us to adopt
a kind of Gupta-Bleuler field quantization.

For later use, the non vanishing inner products between  $\psi_g$, $\psi_s$ and 
$\phi_{Lml}$ and
$(\phi_{Llm})^*$ for $L>0$, read:
 \b
\le\phi_{Llm},\phi_{Llm}\re=1,\
\le\phi_{Llm}^*,\phi_{Llm}^*\re=-1,\
L>0\mbox{ and } 
\le\psi_s,\psi_g\re=1.\e

\section{Gupta-Bleuler triplet}

From now on we shall deal with the minimally coupled field for which we
define the Gupta-Bleuler triplet \cite{bfh,jpg} in order to build a
covariant quantum field. The field equation is given by \b\Box\phi=0.\e
In order to simplify the previous notations, let $K$ be the set of indices for
the positive norm modes, excluding the zero mode:
     $$K=\{(L,l,m)\in\N\times\N\times\Z;\; L\neq0,\, 0\leq l\leq L,\,
-l\leq
     m\leq l\},$$
     and $K'$ the same set including the zero mode:
     $$K'=K\cup\{0\}.$$
As illustrated by (\rf{ncov}), the set spanned by the $\phi_k,\, k\in K$
is not invariant under the action of the de Sitter group. On the other hand, we
obtain an invariant space by adding $\psi_g$. More precisely, let us introduce
the space,
     $$\K=\{c_g\psi_g+\sum_{k\in K}c_k\phi_k;\; c_g,c_k\in\C,\, \sum_{k\in
       K}|c_k|^2<\infty\}.$$
Equipped with the Klein-Gordon-like inner product (\rf{kgie}), $\K$ is a
degenerate inner product space because the above orthogonal basis satisfies to
$$\le\phi_k,\phi_{k'}\re=\delta_{kk'}\ \  \forall k,k'\in K,\hs{10}
 \le\phi_k,\psi_g\re=0\ \  \forall k\in K, \ \mbox{ and }\le\psi_g,\psi_g\re=0.$$
It can be proved by conjugating the action (\rf{ncov}) under the SO(4) subgroup  
that $\K$ is invariant under the natural action of the de Sitter group. As a
consequence, $\K$ carries a unitary representation of the de Sitter group,  this
representation is indecomposable but not irreducible, and the null-norm subspace
$\n=\C\psi_g$ is an uncomplemented invariant subspace. 

Let us recall that the Lagrangian   
  $${\cal L}=\sqrt{|g|}g^{\mu\nu}\partial_\mu\phi\partial_\nu \phi $$
of the free minimally coupled field is invariant when adding to $\phi$ a
constant function. As a consequence, in the ``one-particle sector'' of the 
field, the space of ``global gauge states'' is simply the  invariant 
one dimensional
subspace   $\n=\C\psi_g$. In the following, the space $\K$  is called
the (one-particle) physical space, but {\it stricto sensu} physical states are
defined up to a constant and the space of physical states is $\K/\n$. The
latter is a Hilbert space carrying the unitary irreducible representation of the
de Sitter group $U^{(0)}$.

If one attempts to apply the  canonical
quantization starting from a degenerate space of solutions, then one inevitably
breaks the covariance of the field \cite{dbr}. Hence we must build a non
degenerate invariant space of solutions $\H$ admitting $\K$ as an invariant
subspace. Together with $\n$, the latter are constituent of the so-called Gupta-Bleuler triplet
$\n\subset\K\subset\H$. The construction of $\H$ is worked out as follows.

 We first remark that the modes $\phi_k$ and $\psi_g$ do not form a
complete set of modes. Indeed, the  solution $\psi_s$  does not belong to
$\K$ nor $\K+\K^{*}´$ (where $\K^{*}´$ is
the set of complex conjugates of $\K$): in this sense, it is  not a
superposition of the modes $\phi_k$ and $\psi_g$. One way to prove this is to
note that $\le\psi_s,\psi_g\re=1\neq0$. 

So we need a complete, non-degenerate and invariant inner-product space 
containing $\K$ as a closed subspace. The smallest one fulfilling 
these conditions is the following. Let $\H_{+}$ be the Hilbert space 
spanned by the modes $\phi_{k}$ together with the zero-mode 
$\phi_{0}$:
$$\H_+=\{c_{0}\phi_{0}+´´\sum_{k\in K}c_k\phi_k;\;  \sum_{k\in 
K}|c_k|^2<\infty\}.$$
We now define the total space $\H$
by  $$\H=\H_{+}\oplus\H_{+}^{*}´,$$
which {\it is} invariant, and we denote by $U$ the natural representation of the
de~Sitter group on $\H$ defined by : $U_g\phi(x)=\phi(g^{-1}x)$.  Our 
Gupta-Bleuler triplet is precisely  $\n\subset\K\subset\H$.
The space $\H$ is defined as a direct sum of an Hilbert space and an anti-Hilbert space (a
space with definite negative inner product) which proves that $\H$ is a
Krein space. Note that neither $\H_+$ nor $\H_+^{*}´$ carry a representation of
the de Sitter group, so that the previous decomposition is not 
covariant, although it is O(4)-covariant.
The following family is a pseudo-orthonormal basis for this Krein 
space:
     $$\phi_k,\phi_k^{*}´,(\ k\in K),\;\;\phi_0,\phi_0^{*}´,$$
for which the non-vanishing inner products are
$$\le\phi_k,\phi_k\re=\le\phi_0,\phi_0\re=1\mbox{ and }
\le\phi_k^{*}´,\phi_k^{*}´\re=\le\phi_0^{*}´,\phi_0^{*}´\re=-1.$$

Let us once more insist on the presence of non physical states in 
$\H$. Some of them have  negative norm, but, 
for instance, $\phi_0$ is not a physical state ($\phi_0\not\in\K$) in spite of
the fact that $\le\phi_0,\phi_0\re>0$: the condition of positivity of the inner
product is not a sufficient condition for selecting physical states. Moreover 
some non physical states go to negative frequency states when the 
curvature tends to 0. Nevertheless mean values of observables are computed on 
physical states and no  negative energy  appears.

The space $\H$ is connected to the propagator in the following way.
Since de~Sitter space-time is globally hyperbolic, there
exist two elementary solutions of the field equation, $G^{\rm ret}$ and
$G^{\rm adv}$, which are the unique ones verifying \cite{iun}
$$\Box_xG^{\rm adv}(x,y)=\Box_xG^{\rm ret}(x,y)=-\delta(x,y)$$
and for fixed $y$ the support in $x$ of $G^{\rm adv}(x,y)$ 
(resp. $G^{\rm ret}(x,y)$) lies in the future (resp. the past) of $y$.
The so-called propagator $\tilde G$ is defined by
$$\tilde G=G^{\rm adv}-G^{\rm ret}.$$
 This propagator
is the reproducing kernel with respect to the Klein-Gordon inner product:
$$\phi(\rho,\Omega)=\frac{i}{H^2}\int_{\rho'=0}\!\!\!(-i) \tilde
G\left(\phantom{a^b}\!\!\!(\rho,\Omega),(\rho',\Omega')\right)
\stackrel{\leftrightarrow}{\partial_{\rho'}}\phi(\rho',\Omega')d\Omega',$$
for any  solution $\phi$ of the field equation. Using the above basis of
$\H$, one obtains the following expression for $\tilde G$. 
\begin{eqnarray}
\tilde G(x,x')&=&
\sum_{k\in K}\phi_k^{*}´(x)\phi_k(x')+
\phi_{0}´^{*}´(x)\phi_{0}(x')
-\phi_{0}(x)\phi_{0}^{*}´(x')
-\sum_{k\in K}\phi_k(x)\phi_k^{*}´(x')\nonumber\\ \lbl{wp}
&=&\sum_{k\in K}\phi_k^{*}´(x)\phi_k(x')+
\frac{H}{2\pi}(\psi_s(x')-\psi_s(x))
-\sum_{k\in K}\phi_k(x)\phi_k^{*}´(x').\end{eqnarray}
 
This two-point function $\tilde G$ is linked to $\H$ in the following way.
Since the Riesz representation theorem is valid in Krein spaces, for any continuous
linear form $L$ there exists a unique element $\psi_L\in\H$ such that
$$L(\phi)=\le\psi_L,\phi\re,\ \forall\phi\in\H.$$
This representation theorem (see \cite{dbr} for details) allows one to 
define for any real test function $f$ an element
$p(f)\in\H$ such that
  \b\lbl{defp}\le p(f),\phi\re=\int_{M_H}
f(x)\phi(x)d\mu(x).\e 
This formula defines a $\H$-valued distribution $p$ on $M_H$. In the 
unsmeared form, (\rf{defp}) reads: $\le p(x),\phi\re=\phi(x)$ for any
$\phi\in\H$ . Direct computation with the basis
proves that
\begin{eqnarray}\lbl{p}p(x)&=&
\sum_{k\neq0}\phi_k^{*}´(x)\phi_k
-\sum_{k\neq0}\phi_k(x)\phi_k^{*}´
+\psi_g(x)\psi_s-\psi_s(x)\psi_g,\\\lbl{pdeux}
&=&\sum_{k\neq0}\phi_k^{*}´(x)\phi_k
-\sum_{k\neq0}\phi_k(x)\phi_k^{*}´
+\phi_{0}^{*}´(x)\phi_{0}-\phi_{0}(x)\phi_{0}^{*}´.\end{eqnarray}
Moreover, $-i\tilde G$ is the kernel of $p$, that is to say:
$$\le p(x'),p(x)\re=-i\tilde G(x,x').$$
Note that from (\rf{defp}) one can prove immediately that $p$ commutes with the
action of the de~Sitter group: $U_gp(f)=p(U_gf)$. As a consequence, $\tilde G$ is
invariant:
$$\tilde G(g\cdot x,g\cdot x')=\tilde G(x,x').$$

\section{The Quantum Field}

As explained at the end of section \ref{solutions} the family 
$\phi_{k}$ for $k\in K$ together with $\phi_{0}$ is a complete set of 
mode. As a consequence, we obtain a quantum field through the usual 
formula:
\b\lbl{cq}\varphi(x)=\sum_{k\in K}\phi_k(x)A_k+\phi_{0}(x)A_{0}´´+\sum_{k\in K}
\phi_k^*(x)A\dg_k
+\phi_{0}^{*}(x)A\dg_{0}´´´,\e 
where $A_k$ and $A_k{\dag}$  satisfy the {\bf ccr}:
$$[A_k,A\dg_{k'}]=2\delta_{kk'},\hs{10}
[A_k,A_{k'}]=0,\hs{10}[A\dg_k,A\dg_{k'}]=0,\,k,k'\in K'=K\cup\{0\}.$$
So far this is the usual procedure, except for the insignificant 
factor 2.
However, since the space generated by these modes is not closed under 
the action of de~Sitter group, the usual representation of the 
{\bf ccr} yields a non covariant (although SO(4)-invariant) field.

We now define a new representation of the {\bf ccr} leading to a 
covariant field. Let us first recall that we deal with a  Gupta-Bleuler quantum field.
It is a distribution the values of which are operators on 
 the bosonic Fock space built
on the total space $\H$ (see \cite{m} for the theory of Fock spaces on
Krein spaces). As usual in a Gupta-Bleuler construction, mean values of
observables will be evaluated only with physical states. The physical 
states are the states obtained from the Fock vacuum by creation of one
particle  physical states, which means creation of elements of $\K$.
In a Fock space,  creation and annihilation
operators are defined for arbitrary  states, not only for modes. More 
precisely, let $\uh$
be the Fock space on $\H$, the 
 annihilator  of a solution $\phi$ of the field
equation is defined by: 
$$(a(\phi)\Psi)(x_1,\ldots,x_{n-1})=\sqrt{n} \frac{i}{H^2}\int_{\rho=0}
\phi^{*}´(\rho,\Omega)
\stackrel{\leftrightarrow}{\partial_{\rho}}\Psi((\rho,\Omega),x_1,\ldots,
x_{n-1})d\Omega,$$
for any square-integrable $n$-symmetric function $\Psi$. The creator is
defined as usual by 
$$(a^{\dag}(\phi)\Psi)(x_1,\ldots,x_{n+1})=
\frac{1}{\sqrt{n+1}}
\sum_{i=1}^{n+1}\phi(x_i)\Psi(x_1,\ldots,\check{x_i},\ldots,x_{n+1}).$$

One can easily check  that these operators obey the usual commutation
rules.
\b\lbl{com}[a(\phi),a(\phi')]=0,\hs{10}
[a^{\dag}(\phi),a^{\dag}(\phi')]=0,\hs{10}
[a(\phi),a^{\dag}(\phi')]=\le\phi,\phi'\re,\e
and also
\b\lbl{cov}\uv_ga^{\dag}(\phi)\uv_g^*=a^{\dag}(U_g\phi),\mbox{ and }
\uv_ga(\phi)\uv_g^*=a(U_g\phi),\e
where $U$ is the natural representation of the de Sitter group on $\H$
and $\uv$ its extension to the Fock space.

Let $a_k,\,a_0,\,b_0$ and $b_k$ be the annihilators of the modes $\phi_k,\,
\phi_0,\, \phi_0^{*}´$ and $\phi_k^{*}´$
respectively.
We define 
$$ A_{k}=a_{k}´-b\dg_{k}´,\mbox{ and } A\dg_{k}=a\dg_{k}´-b_{k} 
\mbox{ for } k\in K'= K\cup\{0\}.$$

 The field now reads
\begin{eqnarray}\varphi(x)&=&
\sum_{k\in K}\phi_k(x)a_k+\phi_0(x)a_0-\sum_{k\in K}\phi_k^{*}´(x)b_k
-\phi_0^{*}´(x)b_0
\nonumber\\ &&\hs{50} 
+\sum_{k\in K}\phi_k^{*}´(x)a_k^{\dag}-\sum_{k\in K}
\phi_k(x)b_k^{\dag}
+\phi_0^{*}´(x)a^{\dag}_0-\phi_0(x)b^{\dag}_0,\lbl{field}\end{eqnarray}
The non vanishing commutation relations between the operators in (\rf{field})
are for $k\in K'=K\cup\{0\}$:
\b\lbl{ccr}[a_k,a^{\dag}_k]=1,\hs{10}[b_k,b^{\dag}_k]=-1. \e
Note the  minus sign which follows from the formulas
above and 
$\le\phi_k^{*}´,\phi_k^{*}´\re=-1$.
 Note also that this field is
clearly real as the sum of an operator and its conjugate.

\noindent{\bf Remark} For later use we can rewrite the  field 
$\varphi(x)$ in terms 
of the operators $a_{s}=a(\psi_{s})$ and $a_{g}=a(\psi_{g})$:
\begin{eqnarray}
\varphi(x)&=&
\sum_{k\in K}\phi_k(x)a_k-\sum_{k\in K}\phi_k^{*}´(x)b_k
+\frac{H}{2\pi}a_s+\psi_s(x)a_g\nonumber\\ 
&&\hs{50} +
\sum_{k\in K}\phi_k^{*}´(x)a_k^{\dag}-\sum_{k\in K}\phi_k(x)b_k^{\dag}
+\frac{H}{2\pi}a_s^{\dag}-\psi_s(x)a_g^{\dag}.\lbl{fieldpsi}\end{eqnarray}´´´

We claim that this field is covariant:
$$  \uu_g \varphi(x)\uu^{-1}_g=\varphi(g\cdot x).$$
 This is due to the fact that 
$\H$ is closed under the action of the de~Sitter group, although this 
is not the case for $\H_{+}$.´ In order to prove this statement, we
firstly give a more synthetic expression of the field, directly issued from (\rf{p}) and
(\rf{field}):
\b\lbl{defphi}\varphi(x)=a(p(x))+a^{\dag}(p(x)).\e
It is then straightforward to check that the covariance of $\varphi$ follows
from (\rf{cov}) and the covariance of $p$. Note also that the formula 
(\rf{defphi}) is a coordinate free definition, hence our field does not 
depend on any choice of coordinates.

The causality of the field is also immediate from the values of the commutator
 \b [\varphi(x),\varphi(x')]=2\le p(x),p(x')\re=-2i\tilde G(x,x').\e
We see indeed that the field is causal since $\tilde G$ vanishes when $x$ and $x'$
are space-like separated.

The Gupta-Bleuler vacuum is precisely the Fock vacuum characterized by
$$a_k|0\re=b_k|0\re=0,\ \ \mbox{ for }k\in K'=K\cup\{0\},$$
and it is trivially invariant under the action of de Sitter group. At 
this point, let us emphasize the differences between the usual point 
of view in QFT and ours. The set of modes that we have used in our 
construction is exactly the one used in \cite{a,af} in order to 
obtain the O$(4)$ (and not SO$(1,4)$) invariant vacuum. Nevertheless 
our theory {\it is}  SO$(1,4)$ invariant. This is due to the new 
representation of the {\bf ccr}. Since our field is different from 
the usual one, we think that comparing vacua is misleading.

So we have obtained a quantum field verifying the Wightman axioms.
Some non-physical states are present in the construction, they are really needed in
order to assure the de~Sitter covariance. Now we must define the
observables of the theory. We also have to
verify that non-physical states do not yield any trouble like the appearing of negative energies. 
This is the content of the next section in which we prove that the
stress tensor is an observable and that it is computed directly without any
renormalization. Moreover, expressions like  $\le
n_{k_1}n_{k_2}\ldots|T_{00}|n_{k_1}n_{k_2}\ldots\re$
 are  positive for any physical state $|n_{k_1}n_{k_2}\ldots\re$,
  and no negative energy can be observed.

\section{The stress tensor }

As explained in \cite{dbr}, the global gauge change $\gamma^\lambda$ 
(which would be local if we were dealing with
QED! \cite{bfh}, \cite{jpg})  can be
implemented  in the theory by
$$\gamma^\lambda=\exp\left(-\frac{\pi\lambda}{H}(a^{\dag}_g-a_g)\right),$$
from which one  verifies that
$$\gamma^{-\lambda}\varphi(x)\gamma^\lambda=\varphi(x)+\lambda\id.$$

We now define the (second-quantized) physical space $\uk$ as the space  generated
from the Fock vacuum by creating elements of $\K$, the set of one particle physical
states: $\uk$ is the space generated by the
$(a^{\tiny\dag}_g)^{n_0}(a^{\tiny\dag}_{k_1})^{n_1}\ldots
(a^{\tiny\dag}_{k_l})^{n_l}|0\re$. We call $\uun$ the subspace of $\uk$
orthogonal to $\uk$ \b\lbl{uun}\Psi\in\uun\mbox{ iff }\Psi\in\uk\mbox{ and
}\le\Psi,\Phi\re=0\ \forall\Phi\in\uk.\e 

Note that, when restricted to $\uk$, the operator $a_g$ is the null operator and
that for any physical state $\Psi$, the state $a^{\tiny\dag}_g\Psi\in\uun$. As a
consequence, for any physical state $\Psi$ and any real $\lambda$, the
states $\Psi$ and $\gamma^\lambda\Psi$ are equal up to an element of
$\uun$. This is the motivation for defining elements of $\uun$ as our
second quantized set of global gauge states, and we have obtained our
second-quantized Gupta-Bleuler triplet: $$\uun\subset\uk\subset\uh,$$
which is clearly invariant under the action of the de~Sitter group.
Consistently, two physical
states are said to be physically equivalent when they differ from a global gauge
state and a gauge change transforms a physical state into an
equivalent state.

\noindent{\it Remark  (Quasi-uniqueness of the vacuum):}  The space of the
de~Sitter invariant
 states of $\uh$ is 
$\un$ the space generated from the vacuum by $a^{\tiny\dag}_g$. This space is
an infinite dimensional subspace of $\uun$, hence the Fock vacuum is not
the unique de Sitter invariant state. Nevertheless  one can easily  see that
all these states are physically equivalent to  an element of the one
dimensional space generated by the vacuum state. In this sense we can say
that the vacuum is  unique.

We now have to define the observables of the theory, under the 
condition that a
global gauge change must  not be observed. An observable $A$ is a symmetric
operator on $\uh$ such that, when $\Psi$ and $\Psi'$ are equivalent physical states
(elements of $\uk$ such that $\Psi-\Psi'$ belongs to $\uun$), we must have
$$\le\Psi|A|\Psi\re=\le\Psi'|A|\Psi'\re.$$ 
One can easily verify that the field $\varphi$ is {\it not} an observable. This is
due to the presence of the terms $a_s$ and $a^{\tiny\dag}_s$ 
(\rf{fieldpsi}), these terms disappear
in $\partial_\mu\varphi$ and this is the reason for which the stress tensor
$T_{\mu\nu}$ {\it is} an observable. 
´´

At this point, one can understand the reason why the approach through two-point
functions is not relevant for this field. In fact since the field is 
not an observable, quantities like Wightman or Hadamard functions
$$G(x,x')=\le 0|\varphi(x)\varphi(x')|0\re,\;G^{(1)}(x,x')=
\frac{1}{2}\le 0|\varphi(x)\varphi(x')+\varphi(x')\varphi(x)|0\re,$$
are {\it not gauge invariant}. Hence any definition {\it a priori} of 
such a function in order to obtain a field cannot yield a covariant 
theory. (If one computes $G^{(1)}$ for our field one finds 0, but once 
again, this result has no physical significance).  Actually, there 
 exists no non trivial covariant two-point function of positive 
type, this is nothing but another formulation of Allen's theorem. The 
only two-point function which naturally appears is the commutator, but 
it is not of positive type and it does not allow to select physical states. 
Moreover, the usual classification of vacua is based on two-point 
functions and  our vacuum does not fit this classification. 
We do insist on the fact that it is the field itself which is 
different in our construction and not only the vacuum.

The stress tensor, which in this case is the same as the improved 
stress tensor, is given by \cite{bd}
       $$T_{\mu\nu}=\partial_\mu\phi\partial_\nu\phi-\frac{1}{2}
          g_{\mu\nu}g^{\rho\sigma} \partial_\rho\phi\partial_\sigma\phi.$$ 
Let us consider the excited physical state 
$$|\vec k\re=|k_1^{n_1}\ldots k_j^{n_j}\re=\frac{1}{\sqrt{n_1!\ldots n_j!}}
\left(a_{k_1}^{\dag}\right)^{n_1}
\ldots\left(a_{k_j}^{\dag}\right)^{n_j}|0\re.$$
In order to compute 
$\le \vec k|T_{\mu\nu}(x)|\vec k\re$, we
begin with $\le \vec k|\partial_\mu\varphi(x)\partial_\nu\varphi(x)|\vec k\re$.
As mentioned before, the terms containing $a_s$ and $a^{\tiny\dag}_s$ disappear in the
derivation. Moreover $a_g$ and $a^{\tiny\dag}_g$ commute with all the remaining
operators including themselves, and so the corresponding terms vanish in the
computation. We then obtain
$$\le \vec k|\partial_\mu\varphi(x)\partial_\nu\varphi(x)|\vec k\re=
\sum_{k\in K}\partial_\mu\phi_k(x)\partial_\nu\phi^{*}´_k(x)
-\partial_\mu\phi_k^{*}´(x)\partial_\nu\phi_k(x)
+2\sum_{i=1}^ln_i
\Re\left(\partial_\mu\phi_{k_i}^{*}´(x)\partial_\nu\phi_{k_i}(x)\right).$$
The first and third terms are those one obtains in the usual computation, the
first one carries infinite terms which have to be renormalized in the usual theory.
The unusual second term, with the minus sign, is due to the presence of $b_k$ and $b^{\tiny\dag}_k$ in
the field. Thanks to this term, there is no need to renormalize the stress tensor
because one obtains immediately that
$$\le \vec k|\partial_\mu\varphi(x)\partial_\mu\varphi(x)|\vec k\re
=2\sum_{i=1}^ln_i
\partial_\mu\phi_{k_i}^{*}´(x)\partial_\mu\phi_{k_i}(x).$$
A direct consequence of this formula is the positivity of the energy, more
precisely, one can see at once  that
$$\le\vec k|T_{00}|\vec k\re\geq0,$$
 for any physical state $|\vec k\re$ 
 and that this quantity vanishes if and only if $|\vec k\re=|0\re$. 
 This is not in contradiction with the existence of other states for 
 which the energy can be negative, but of course these states are not 
 physical. 
 One can see that the non-physical states do not play any role for the 
 free field. In the interacting case, the situation would be different.
 One can then expect the appearing of some virtual particles like for 
 QED in presence of charges 
 (see \cite{mandl} for instance).
 
\noindent{\it Remarks on the renormalization}: The present 
renormalization  is fully covariant and has nothing to do neither with 
the choice of modes nor with the presence of zero modes. It is totally 
different from  other existing renormalization procedures.
This is due to
 $\le\partial_\mu p(x),\partial_\mu p(x)\re=0$ which 
implies that 
$$[a(p(x)),a\dg(p(x))]=0.$$
Moreover, this renormalization eliminates infra-red as well as 
ultra-violet divergence. Actually, both divergences are carried by the 
Hadamard function $G^{(1)}$ and the latter vanishes here. 
Finally,       this renormalization fulfills the so-called Wald axioms. 
\begin{enumerate}
\item The stress tensor is covariant and causal since the field is. 
\item The computation above shows that it furnishes the usual ({\it i.e.} 
formal) results for physical states. \item The corner stone of the above 
computation is the following:
$$[b_k,b\dg_k]=-1,$$
which implies that
$$a_ka\dg_k+a\dg_ka_k+b_kb\dg_k+b\dg_kb_k=2a\dg_ka_k+
2b\dg_kb_k.$$
One can see that this is equivalent to reordering when applied to 
physical states (on which $b_{k}$ vanishes).\end{enumerate}

In conclusion, we have introduced  auxiliary states (states which do not belong to
$\K$) for constructing a covariant quantization of the massless minimally
coupled scalar field. But the effect of these auxiliary states appears in the
physics of the problem  by allowing an automatic renormalization of the stress
tensor, and, once again, the auxiliary states do not yield any measurable negative
energy.

\section{Back to the massive field}

As explained in the above, the crucial point about the minimally coupled field 
is the fact that there does not exist a covariant decomposition
$$\H=\H_+\oplus\H_-,$$
where $\H_+$ (resp. $\H_-$) is a Hilbert space (resp. anti-Hilbert space).
This was the reason for which our space of states contains negative frequency
solutions. It is not the case for the scalar massive field for which such a
decomposition exists, where $\H_+$ is the usual physical states space 
and $\H_{-}=\H_{+}^{*}$´´´.

Nevertheless one can define a $\H$-valued distribution $p$  as in (\rf{defp})
and  a field $\varphi$ as in (\rf{defphi}), except that the above decomposition
of $\H$ yields a covariant decomposition $p=p_++p_-$ with
$p_{\varepsilon}(x)\in\H_{\varepsilon}$ and a decomposition of the field into two
parts $$\varphi=\varphi_++\varphi_-.$$
The positive frequency part $\varphi_+$, written in terms of annihilators and creators
is exactly the usual field. Moreover, for $\Psi$ and $\Psi'$ physical states, we
have
$$\le\Psi|\varphi(x)|\Psi'\re =\le\Psi|\varphi_+(x)|\Psi'\re.$$
However, this does not mean that $\varphi$ and $\varphi_+$ are the same object,
as operators they are different and quantities like
$$\le\Psi|\varphi(x)\varphi(x')|\Psi'\re\mbox{ and }
\le\Psi|\varphi_+(x)\varphi_+(x')|\Psi'\re$$
{\it are} different. In particular, the energy-momentum tensor computation 
presented in the previous section 
can be easily transposed to the present massive scalar field with the following
issue. {\it Using the Gupta-Bleuler quantization for the massive
field provides an automatic and covariant renormalization of the vacuum energy
divergence.} Note that the conformal massless case is in some sense a 
particular case of the massive one and our construction supplies a field 
which is covariant and conformally covariant in a strong sense. As a 
consequence, the trace anomaly does not appear.
 This is not surprising because, after all, trace anomaly can appear only by 
breaking the conformal invariance.

In a future work \cite{gr}, we show that this quantization 
explains the appearance of negative frequency terms in the flat limit 
of de~Sitter space-time.

As a final remark, relevant to both massive and massless 
case,           let us discuss the Bogolioubov transformations in our
Gupta-Bleuler framework. First of all, let us point out that from  our point of
view, there is a unique vacuum, the (Krein-)Fock vacuum, which is invariant and
normalizable. This does not mean that the Bogolioubov transformations are no longer
valid in Gupta-Bleuler quantization. Any element like
$\tilde\phi_k=A_k\phi_k+B_k\phi_k^{*}´$ belongs to $\H$, and a Bogolioubov
transformation is just a change of physical states. The new
physical space  is $\tilde\H_+=\mbox{\rm span}(\tilde\phi_k)$, for which there corresponds
a new $\tilde\varphi_+$. If one wants to characterize the new physical space by
some two-point function, one can compute
$$\tilde G_{+}^{(1)}=\le0|\tilde\varphi_+(x)\tilde\varphi_+(x')
+\tilde\varphi_+(x')\tilde\varphi_+(x)|0\re.$$
This gives exactly the same family of function as the expression (2.14) of \cite{a}.

\section{Conclusion and outlook}

Any consistent approach to quantization of fields in de Sitter space-time has
to deal with  the
negative-energy problem from a minkowskian point of view (see (\rf{contr})). This
problem of ``negative-frequency'' modes from a curved space-time point of view is
also present in the manipulation of the zero-modes.

Different ways to go round this problem have been proposed in order to reach a
point in the theory where only positive-energy states are taken into account: 
restriction to a subgroup \cite{a}, analyticity
constraint (massive case) \cite{bgm}, modification of the vacuum definition
\cite{af,kg}. 

Another difficulty appears when dealing with fields involving a gauge invariance.
The Gupta-Bleuler formalism has been created in order to manage both covariance
and gauge invariance in quantum electrodynamics. It is not surprising that an
analogous construction accomplishes the same task for the minimally-coupled field on
de~Sitter space-time.  We have here  presented a new proposal which is a
continuation of \cite{dbr}. The guideline is the
covariance of the full theory under the full SO$_{\rm o}(1,4)$ in the spirit of the
Wightman-G\"{a}rding approach. The fact that our Krein-Gupta-Bleuler quantization
gives the correct sign for the energy pleads in favor of an extension of this
method to  the massless spin-2 case on de Sitter space-time. The
corresponding field is indeed built up from two copies of minimally coupled
fields \cite{ta} and we hope that the present paper will open the way to
a satisfactory covariant quantization of this field.

\vskip15pt \noindent {\it Acknowledgments}: The authors would like to
thank  J. Iliopoulos and D. Polarski for helpful discussion, and the 
referees for constructive comments.

\end{document}